\documentclass[10pt]{amsart}
\usepackage{graphicx}
\usepackage{amssymb}
\usepackage{amsfonts}
\usepackage{amsmath}
\usepackage{amsthm}

\usepackage[all]{xy}
\newtheorem{theorem}{Theorem}[section]

\theoremstyle{definition}
\newtheorem{definition}[theorem]{Definition}

\newtheorem{remark}[theorem]{Remark}

\numberwithin{equation}{section}
 \theoremstyle{plain}    
 
 \theoremstyle{remark}    
 \newtheorem*{acknowledgement*}{Acknowledgement} 

\newcommand{\cF}{\mathcal{F}}
\newcommand{\cG}{\mathcal{G}}

\newcommand{\cJ}{\mathcal{J}}

\newcommand{\cP}{\mathcal{P}}

\newcommand{\cS}{\mathcal{S}}
\newcommand{\cT}{\mathcal{T}}

\newcommand{\cX}{\mathcal{X}}

\newcommand{\te}{{\theta}}
\newcommand{\Te}{{\Theta}}

\newcommand{\Om}{{\Omega}}
\newcommand{\om}{{\omega}}
\newcommand{\ve}{{\varepsilon}}
\newcommand{\del}{{\delta}}
\newcommand{\Del}{{\Delta}}
\newcommand{\gam}{{\gamma}}
\newcommand{\Gam}{{\Gamma}}
\newcommand{\vf}{{\varphi}}

\newcommand{\up}{{\upsilon}}

\newcommand{\sig}{{\sigma}}
\newcommand{\al}{{\alpha}}
\newcommand{\be}{{\beta}}
\newcommand{\ka}{{\kappa}}
\newcommand{\la}{{\lambda}}

\newcommand{\bbE}{{\mathbb E}}

\newcommand{\bbN}{{\mathbb N}}
\newcommand{\bbP}{{\mathbb P}}
\newcommand{\bbR}{{\mathbb R}}

\newcommand{\bbI}{{\mathbb I}}

\newcommand{\bfz}{{\bf z}}
\newcommand{\bfu}{{\bf u}}

\swapnumbers
\theoremstyle{plain}

\numberwithin{equation}{section}

\begin{document}
\title[]{Dynkin games and Israeli options}%
 \vskip 0.1cm
 \author{ Yuri Kifer\\
\vskip 0.1cm
Institute of Mathematics\\
Hebrew University\\
Jerusalem, Israel}%
\address{
 Institute of Mathematics, The Hebrew University, Jerusalem 91904, Israel}%
\email{  kifer@math.huji.ac.il}%

\thanks{Partially supported by the ISF grant no. 82/10}

 \date{\today}
\begin{abstract}\noindent
 We start by briefly surveying research on optimal stopping games since 
 their introduction
 by E.B. Dynkin more than 40 years ago. Recent renewed interest to Dynkin's 
 games is due, in particular, to the study of Israeli (game) options introduced in
 2000. We discuss the work on these options and related derivative securities
 for the last decade. Among various results on game options we consider error
 estimates for their discrete approximations, swing game options, game options
 in markets with transaction costs and other questions.
\end{abstract}
\maketitle
\markboth{ Y. Kifer}{Dynkin games and Israeli options}
\renewcommand{\theequation}{\arabic{section}.\arabic{equation}}
\pagenumbering{arabic}

\section{Introduction}\label{sec1}

Optimal stopping games were introduced in 1969 by E.B. Dynkin in \cite{Dy}
as an extension of the optimal stopping problem which was already actively
 studied since 1950ies. Optimal stopping and, in particular, its game version 
 was often discussed on Dynkin's undergraduate seminar at Moscow State
 University in the end of 1960ies which resulted in papers \cite{Fr},
  \cite{GZ}, \cite{Ki1} and \cite{Ki2}.
  
 The original setup of optimal stopping games consisted of a probability
 space $(\Om,\cF,\bbP)$, of a filtration of $\sig$-algebras $\{\cF_t\}$,
 $\cF_t\subset\cF$ with either $t\in\bbN=\{ 0,1,2,...\}$ (discrete time case)
 or $t\in\bbR_+=[0,\infty)$ (continuous time case), of $\{\cF_t\}$-adapted 
 payoff process $\{ X_t\}$ and of a pair of $\{\cF_t\}$-adapted 0--1 valued
 "permission" processes $\vf_t^{(i)},\, i=1,2$ such that the player $i$ is
 allowed to stop the game at time $t$ if and only if $\vf_t^{(i)}=1$. If
 the game is stopped at time $t$ then the first player pays to the 2nd one
 the sum $X_t$. Clearly, if $\vf_t^{1}\equiv 1$ and $ \vf_t^{(2)}\equiv 0$ we
 arrive back at the usual optimal stopping problem. Observe that in the one
 player optimal stopping problem the goal is maximization of the payoff and
 the corresponding supremum always exists (may be infinite), so
 only optimal or almost optimal stopping times remain to be found while in the
 game version already existence of the game value is the question which
 should be resolved first and only then we can look for optimal (saddle
 point) or almost optimal stopping times of the players.
 
  Few years later J. Neveu suggested in \cite{Ne} a very useful 
  generalization of the above setup which turned out to be more convenient 
  both for further study and for applications. Namely, now the "permission"
  processes were dropped off and the players could stop whenever they want
  but instead two payoff adapted processes $X_t\geq Y_t$ were introduced.
  It was prescribed that if the 1st player stops the game at time $s$ and the
  2nd one at time $t$ then the former pays to the latter the amount $X_s$
  or $Y_t$ if $s\leq t$ or $s>t$, respectively. If desired we can have 
  virtual "permission" processes within this setup not by direct regulations
  but by a "market economy" tools. Namely, in order to accomplish this it
  suffices to prescribe very high payment
  $X_t$ or a very low (may be negative) payment $Y_t$ where we "forbid"
  to stop the game by the 1st player or by the 2nd one, respectively.
  
  We observe that from a bit different perspective differential games with 
  stopping times were studied in 1970ies in a series of papers (see \cite{BF1},
  \cite{BF2} and references there). Game versions of optimal stopping of a
  Markov process and of a diffusion were considered in \cite{El} and \cite{Bis},
  respectively. It seems that the term "Dynkin game" appeared first in 
  \cite{ANLM}.
  
  Israeli or game options were introduced first in \cite{Ki3} though some
  special callable derivative security LION was discussed before in \cite{MS}
  in a kind of game framework without any rigorous justification. An option 
  or a contingent claim is certain contract and an American option enables
  its buyer (holder) to exercise it at any time up to the maturity. A game
  option gives additionally the right to the option seller (writer, issuer)
  to cancell it early paying for this a prescribed penalty. The rational
  behind this provision comes from an idea that essentially any contract
  stipulates conditions for a way out so that the financial market should 
  not be an exception.
  
  The classical approach to pricing of options is based on hedging arguments. 
  Namely, the price is defined as a minimal initial amount of a self-financing 
  portfolio which can provide protection (hedging) against any exercising
  strategy of the option's buyer. So, somehow heuristically, this leads to 
  the infimum over the seller's strategies and to the supremum over the
  buyer's strategies, i.e. we arrive at a game type $\inf\sup$ representation
  which still should be rigorously justified.
  
  The structure of this paper is the following. In the next section we briefly
  survey main results concerning Dynkin's games. In Section 3 we discuss the
  up to date research about game options and related derivative securities.
  In Sections 4 and 5 we exhibit more special results concerning discrete
  approximations of game options and game options in markets with transaction 
  costs, respectively.

  \section{Dynkin games}\label{sec2}

The general modern setup for a Dynkin's game consists of a probability space 
$(\Om,\cF,\bbP)$, of a right continuous filtration of complete $\sig$-algebras
$\{\cF_t\}$ and of three $\{\cF_t\}$-adapted stochastic processes $X_t,\, Y_t$
and $Z_t$ so that when the 1st player stops the game at time $s$ and the
2nd one stops at time $t$ then the former pays to the latter the amount
\begin{equation}\label{2.1}
H(s,t)=X_s\bbI_{s<t}+Y_t\bbI_{s>t}+Z_t\bbI_{s=t}
\end{equation}
where $\bbI_\Gam=1$ if an event $\Gam$ occurs and $=0$, otherwise. We allow
the time $t$ to run either along nonnegative integers $\bbN$ or along 
nonnegative reals $\bbR_+$ up to some horizon $T\leq\infty$ when the game is 
stopped and the 1st player pays to the 2nd one the amount
\[
H(T,T)=X_T=Y_T=Z_T
\]
where in the case $T=\infty$ we assume that
\begin{equation}\label{2.2}
0=X_\infty=\lim_{t\to\infty}X_t=Y_\infty=\lim_{t\to\infty}Y_t=
Z_\infty=\lim_{t\to\infty}Z_t.
\end{equation}
In the continuous time case, i.e. when $t$ runs over $\bbR_+$, the processes
$X_t,Y_t$ and $Z_t$ are supposed to be right continuous.

Next, assume that for any $t\in[0,T]$,
\begin{equation}\label{2.3}
Y_t\leq Z_t\leq X_t\quad\bbP-\mbox{almost surely (a.s)}\quad\mbox{and}
\end{equation}
\begin{equation}\label{2.4}
\bbE\sup_{0\leq t\leq T}(|Y_t|+|Z_t|+|X_t|)<\infty.
\end{equation}
Denote by $\cT_{s,t}.\, s\leq t$ the collection of all stopping times $\tau$
with values between $s$ and $t$ (i.e. nonnegative random variables such that
$\{\tau\leq u\}\in\cF_u$ for all $u$). Introduce the upper and the lower
values of the game starting at time $t\leq T$ by
\begin{equation}\label{2.5}
\bar V_t=ess\inf_{\sig\in\cT_{tT}}ess\sup_{\tau\in\cT_{tT}}
\bbE\big(H(\sig,\tau)| \cF_t\big)\,\,\mbox{and}
\end{equation}
\begin{equation}\label{2.6}
\underbar V_t=ess\sup_{\tau\in\cT_{tT}}ess\inf_{\sig\in\cT_{tT}}
\bbE\big(H(\sig,\tau)| \cF_t\big).
\end{equation}
It turns out that we can choose these processes $\{ \bar V_t\}$ and
$\{ \underbar V_t\}$ to be right upper semicontinuous which is a sufficient
regularity in order to proceed here. 

\begin{theorem}\label{thm2.1}
Under the above conditions $V_\tau\overset{def}{=}\bar V_\tau=
\underbar V_\tau$ a.s. for any stopping time $\tau\in\cT_{0T}$ and,
 in particular, the Dynkin's game has a value
\begin{equation}\label{2.7}
V=V_0=\bar V_0=\underbar V_0.
\end{equation}
Furthermore, for any $\ve>0$ the stopping times
\begin{equation}\label{2.8}
\sig_\ve=\inf\{ t\leq T:\, V_t\geq X_t-\ve\}\,\,\mbox{and}\,\,
\tau_\ve=\inf\{ t\leq T:\, V_t\leq Y_t+\ve\}
\end{equation}
are $\ve$-optimal, i.e. for any $\sig,\tau\in\cT_{0T}$,
\begin{equation}\label{2.9}
\bbE\big(H(\sig_\ve,\tau))-\ve\leq\bbE\big( H(\sig_\ve,\tau_\ve)\big)\leq
\bbE\big(H(\sig,\tau_\ve)\big)+\ve.
\end{equation}
Under additional regularity conditions (say, $X_t,\, Y_t,\, Z_t$ are continuous
stochastic processes) the inequality (\ref{2.9}) remains true for $\ve=0$ with
some $\sig_0,\,\tau_0$, i.e. there exists a saddle point for the Dynkin's game
above. In the discrete time case we have also the following backward recursive 
(dynamical programming) relation
\begin{equation}\label{2.10}
V_n=\min\big(X_n,\,\max(Y_n,\bbE(V_{n+1}|\cF_n))\big).
\end{equation}
\end{theorem}

The theorem above follows from \cite{Ne},\,\cite{Oh1},\,\cite{Oh2} and 
\cite{Mo1} in the discrete time case and from \cite{LM}, \cite{St2} and
\cite{Oh4} in the continuous time case. Observe that 
(\ref{2.5})--(\ref{2.7}) and (\ref{2.9}) imply
\[
|V_0-\bbE\big( H(\sig_\ve,\tau_\ve)\big)|\leq\ve.
\]

If the condition (\ref{2.3}) does not hold true then the above game value 
may not exist (i.e. $\bar V_0>\underbar V_0$) if the players are restricted to 
usual (pure) stopping times and to have the game value they should be allowed
to use randomized stopping times (see \cite{Ya1},\,\cite{RSV},\,\cite{TV},\,
\cite{NRS} and \cite{LS}). Other results on Dynkin's games leading to
randomized stopping times can be found in \cite{Dom1}, \cite{Dom2}, 
\cite{Fe1} and \cite{Fe2}.

\begin{remark}\label{rem2.2} We observe that randomized stopping times used
in the above mentioned papers in order to obtain Dynkin's game value without
the condition (\ref{2.3}) look somewhat different from randomized stopping
times we employ in Section \ref{sec5} in order to study game options in 
markets with
transaction costs. Namely, the above papers deal with randomized stopping 
times having (in the discrete time case) the form $\la(p)=\min\{ n\geq 0:\,
A_n\leq p_n\}$ where $p=(p_0,p_1,...)$ is an adapted to the filtration
$\{\cF_n\}$ process with $p_n\in[0,1]$ for all $n$ and $A_0,A_1,A_2,...$ is
a sequence of independent identically uniformly distributed on $[0,1]$ 
random variables independent of payoff processes. Sometimes, it is assumed
additionally (see \cite{RSV}) that $A_n$ is $\cF_{n+1}$-measurable and 
independent of $\cF_n$. If $W=(W_0,W_1,W_2,...)$ is an adapted stochastic 
process then we can write
\[
W_{\la(p)}=\sum_{n=0}^\infty\psi_nW_n\,\,\mbox{where}\,\,\psi_n=
\bbI_{\{ A_n\leq p_n\}}\prod_{j=0}^{n-1}\bbI_{\{ A_j> p_j\}}.
\]
On the other hand, randomized stopping times employed in Section 5 are 
determined
 by an adapted nonnegative sequence $\chi=(\chi_0,\chi_1,\chi_2,...)$ such
 that $\sum_{j=0}^\infty\chi_j=1$ and for an adapted stochastic process $W$
 as above we write $W_\chi=\sum_{n=0}^\infty\chi_nW_n$. Here $\{\chi_n\}$ is
 an adapted sequence but not necessarily indicators of events while the above
  sequence $\{\psi_n\}$ is not adapted (unless the filtration is properly
  enlarged) and it consists of indicators of events. Still, with respect to
  the enlarged filtration $\la(p)$ is a usual (pure) stopping time while
  randomized stopping times of Section \ref{sec5} look rather differently.
  Nevertheless, it turns out that these two approaches to randomized
  stopping times are essentially equivalent if $\prod_{n\geq 0}(1-p_n)=0$
  (see \cite{SV2} in the discrete time case and the corresponding discussion
   in \cite{TV} for the continuous time case).
  \end{remark}

Among other works on Dynkin's games we can mention results on non-zero-sum
games (see \cite{Na},\,\cite{NS},\,\cite{Mo2} and \cite{Oh3}), Dynkin's
games with asymmetric information (see \cite{LM}), more than 2
 person optimal stopping games (see \cite{YNK}, \cite{Mo3} and \cite{SV}),
  optimal 
 stopping games driven by Markov processes (see \cite{Fr},\,\cite{St1},\,
 \cite{El} and \cite{EP}), Dynkin's games via backward stochastic 
 differential equations with reflection (see \cite{CK},\,\cite{HL} and
 \cite{HW}) and via Dirichlet forms (see \cite{FT}), as well as some other
 results on Dynkin's and similar games (see \cite{Ya2}, \cite{Oh5}, \cite{Oh6},
 \cite{Ka}, \cite{KW}, \cite{Pe}, \cite{BK2}, \cite{BK3}, \cite{BK4}, 
 \cite{EV} and \cite{Al0}).

\section{Game options and their shortfall risk}\label{sec3}

A game (Israeli) option (or contingent claim) studied in \cite{Ki3} is a 
contract between a writer and a holder at time $t=0$ such that both have the
 right to exercise at any stopping 
time before the expiry date $T$. If the holder exercises at time $t$ he or she
receives the amount $Y_t\geq 0$ from the writer and if the writer exercises at 
time $t$ before the holder
he must pay to the holder the amount $X_t\geq Y_t$ so that 
$\del_t=X_t-Y_t$
 is viewed as a penalty imposed on the writer for cancellation of the contract.
 If both exercise at the same time $t$ then the holder may claim $Y_t$ and if
 neither have exercised until the expiry time $T$ then the holder may claim 
 the amount $Y_T$. In short, if the writer will exercise at a stopping time
 $\sig\leq T$ and the holder at a stopping time $\tau\leq T$ then the former
 pays to the latter the amount $H(\sig,\tau)$ where
 \begin{equation}\label{3.1}
 H(s,t)=X_s\bbI_{s<t}+Y_t\bbI_{t\leq s}.
 \end{equation}
 We consider such game options in a standard securities 
 market consisting of a nonrandom component $b_t$ representing the value of
a savings account at time $t$ with an interest rate $r$ and of a random
component $S_t$ representing the stock price at time $t$. As
usual, we view $S_t,t>0$ as a stochastic process on a complete probability
space $(\Omega,\mathcal{F},P)$ and we assume that it generates a
right continuous filtration $\{\mathcal{F}_{t}\}$ and that the
payoff processes $X_t$ and $Y_t$ are right continuous processes adapted 
to this filtration and satisfying the integrability conditions (\ref{2.4}).
 
The classical approach suggests that valuation of options should be based
on the notions of a self-financing portfolio and on hedging. We start with
a portfolio strategy $\pi=\{\pi_t\}_{0\leq t\leq T}$ which is a collection
of pairs $\pi_t=(\be_t,\gam_t)$ so that the portfolio value $W_t$ at time $t$
equals 
\[
W^\pi_t=\be_tb_t+\gam_t S_t
\]
where the process $(\be_t,\gam_t),\, 0\leq t\leq T$ is supposed to be 
predictable in the discrete time case and progressively measurable in the
continuous time case. A portfolio strategy $\pi$ is called self-financing
if all changes in the portfolio value are due to capital gains or losses
but not to withdrawal or infusion of funds. This can be expressed by the
relations (see \cite{Sh}),
\[
b_{t-1}(\be_t-\be_{t-1})+S_{t-1}(\gam_t-\gam_{t-1})=0\,\,\mbox{for}\,\,
t=1,2,...,T
\]
in the discrete time case and 
\[
W^\pi_t=W^\pi_0+\int_0^t\be_udb_u+\int_0^t\gam_udS_u
\]
in the continuous time case. We assume also in the continuous time case
that with probability one
\[
\int_0^T|b_t\be_t|dt<\infty\,\,\mbox{and}\,\,\int_0^T(\gam_tS_t)^2dt<\infty.
\]
A pair $(\sig,\pi)$ of a stopping time 
$\sig\leq T$ and a self-financing portfolio strategy $\pi$ is called 
a hedge (against the game contingent claim) if $W^\pi_{\sig\wedge t}
\geq H(\sig,t)$ with probability one for any $t\in[0,T]$. Now the fair
price of the game option is defined as the infimum of capitals $x$ 
for which there exists a hedge $(\sig,\pi)$ with $W^\pi_0=x$. In a
complete market (i.e. having a unique martingale measure) this is a widely 
acceptable fair price of the option while in an incomplete market or
in a market with transaction costs this definition leads to what is
known as superhedging (see \cite{Sh}).

 Two popular models of complete markets were considered
 in \cite{Ki3} for pricing of game options. First, the discrete time CRR 
 binomial model (see \cite{CRR}) was
 treated there where the stock price $S_k$ at time $k$ is equal to
 \begin{equation}\label{3.2}
 S_k=S_0\prod_{j=1}^k(1+\rho_j),\,\,\, S_0>0
 \end{equation}
 where $\rho_j,\, j=1,2,...$ are independent identically distributed 
 (i.i.d.) random variables such that $\rho_j=
 b>0$ with probability $p>0$ and $\rho_j=a<0,\, a>-1$ with probability
 $q=1-p>0$. Secondly, \cite{Ki3} deals with the continuous time Black-Scholes
 (BS) market model where the stock price $S_t$ at time $t$ is given by the 
 geometric Brownian motion
 \begin{equation}\label{3.3}
 S_t=S_0\exp\big( (\al-\ka^2/2)t+\ka B_t\big),\,\,\, S_0>0
 \end{equation}
 where $\{ B_t\}_{t\geq 0}$ is the standard one-dimensional continuous in
 time Brownian motion (Wiener process) starting at zero and $\ka>0$,
 $\al\in(-\infty,\infty)$ are some parameters. In 
 addition to the stock which is a risky security  the market includes in both 
 cases also a savings account with a deterministic growth given by the
 formulas
 \begin{equation}\label{3.4}
 b_n=(1+r)^nb_0\,\,\,\,\,\mbox{and}\,\,\,\,\, b_t=b_0e^{rt},\,\, b_0,r>0
 \end{equation}
 in the CRR model (where we assume in addition that $r<b$) and in the BS model,
  respectively.
 
 Recall (see \cite{Sh}) that a probability measure describing the evolution
 of a stock price in a stochastic financial market is called martingale 
 (risk-neutral) if the discounted stock prices ($(1+r)^{-k}S_k$ in the CRR
 model and $e^{-rt}S_t$ in the BS model) become martingales.
  Relying on the above hedging arguments the following result was proved
   in \cite{Ki3}. 
 \begin{theorem}\label{thm3.1}  
  The fair price $V$ of the game option is given by the formulas 
 \begin{equation}\label{3.5}
 V=\min_{\sig\in\cT_{0T}}\max_{\tau\in\cT_{0T}}E\big((1+r)^{-\sig\wedge\tau}
 H(\sig,\tau)\big)
 \end{equation}
 in the CRR market (with usual notations $a\wedge b=\min(a,b)$, $a\vee b=
 \max(a,b))$ and
 \begin{equation}\label{3.6}
 V=\inf_{\sig\in\cT_{0T}}\sup_{\tau\in\cT_{0T}}
 E\big(e^{-r\sig\wedge\tau}H(\sig,\tau)\big)
 \end{equation}
 in the BS market where the expectations are taken with respect to
 the corresponding martingale probabilities, which are
 uniquely defined since these markets are known to be complete (see \cite{Sh}), $T$ is the expiry time and 
 $\cT_{st}$ is the space of corresponding stopping times with values between
 $s$ and $t$ taking into account that
  in the CRR model $\sig$ and $\tau$ are allowed to take only
 integer values.
 \end{theorem}
 
 Observe, that the formulas (\ref{3.5}) and (\ref{3.6}) 
 represent also the values of corresponding Dynkin's (optimal stopping) games 
 with payoffs $(1+r)^{-\sig\wedge\tau}H(\sig,\tau)$ and $e^{-r\sig\wedge\tau}
 H(\sig,\tau)$, respectively, when the first and the second players stop the 
 game at stopping times $\sig$ and $\tau$, respectively. 
 The continuous time BS model is generally considered as a better description
 of the evolution of real stocks, in particular, since the CRR model allows
 only two possible values $(1+b)S_k$ and $(1+a)S_k$ for the stock price 
 $S_{k+1}$ at time $k+1$ given its price $S_k$ at time $k$. The main advantage
 of the CRR model is its simplicity and the possibility of easier computations
 of the value $V$ in (\ref{3.5}), in particular, by means of the dynamical 
 programming recursive relations (see \cite{Ki3}),
 \begin{eqnarray}\label{3.7}
 &V=V_{0,N},\,\,\,\, V_{N,N}=(1+r)^{-N}Y_N,\,\,\,\mbox{and}\\
 &V_{k,N}=\min\bigg((1+r)^{-k}X_k,\,\max\big((1+r)^{-k}Y_k,\,
  E(V_{k+1,N}|\cF_k)\big)\bigg)\nonumber 
 \end{eqnarray}
 where a positive integer $N$ is an expiry time and $\{\cF_k\}_{k\geq 0}$
 is the corresponding filtration of $\sig$-algebras. By this reason it makes
 sense to study approximations of the BS model by CRR models which we describe
 in the next section.
 
 Recently it became popular to employ game options as a framework for the study
 of convertible (callable) bonds (see \cite{KaK2}, \cite{SS}, \cite{GK}, 
 \cite{SPS}, \cite{KvS},
 \cite{BCJR4} and \cite{YS2}). A holder of such bond either does nothing
 or decides to convert it into a predetermined number of stocks which can
  be considered as a cash payment depending on the current stock price,
  especially, in a market without transaction costs. On the other hand,
   the firm which issued this callable convertible bond  may redeem it any
   time at a call price or force its conversion into stocks, and so this 
   situation can be treated within the setup of game options.
   
   Several papers deal with computation of the fair price of game options in
   special situations when the underlying stock price evolves according to
   a Markov process which usually, as in the BS model, turns out to be the
   geometric Brownian motion and when the payoffs depend only on the current
   stock price, usually just for the put and call options payoffs arriving at
   a study of the free boundary problem with buyer's and seller's exercise
   boundaries (see \cite{Ky}, \cite{KS}, \cite{KK}, \cite{KKS}, \cite{SuS},
   \cite{SSY} and \cite{YYZ1}). For other callable derivative securities which
   were studied within the game options framework and its generalizations we 
   refer the reader to \cite{BCJR1}, \cite{BCJR3}, \cite{YS1}, \cite{SSW},
   \cite{Al} and \cite{YYZ2}.

In real market conditions an investor (seller) may not be willing for
various reasons to tie in a hedging portfolio the full initial
capital required for a (perfect) hedge. In this case the seller is
ready to accept a risk that his portfolio value at an excercise time
may be less than his obligation to pay and he will need additional
funds to fullfil the contract. Thus a portfolio shortfall comes into
the picture and it is important to estimate the corresponding risk.
We consider here a certain type of risk called the
shortfall risk which was defined for game options in \cite{DK1} by
\begin{equation}\label{3.8}
R(x)=\inf_{(\pi,\sigma)}R(\pi,\sigma)\,\,\mbox{where}\,\,
R(\pi,\sigma)=\sup_{\tau}\bbE\big((H(\sigma,\tau)-b_0W^\pi_{\sigma\wedge\tau})^+/
b_{\sigma\wedge\tau}\big)
\end{equation}
where the supremum is taken over all self-financing portfolio strategies $\pi$
with an initial capital $x$
and both in infimum and in supremum the stopping times $\sig$ and $\tau$ do
not exceed the option expiration date (horizon) $T$. It was shown in \cite{DK1}
that in the discrete time case both the shortfall risk and the corresponding
minimizing portfolio strategies and stopping times could be obtained by means
of a backward induction (dynamical programming) algorithm. In the continuous
time case the situation is more complicated. For the shortfall risk in the
American options case \cite{Mu} obtained existence of minimizing strategies
relying on some convex analysis argument which are not available in the
game options case, and so existence of minimizing portfolio strategies and
 stopping times in (\ref{3.8}) remains an open question.
 
 The papers \cite{DIK} and \cite{IK1} deal with, so called, swing game options
 which are, in fact, multiple exercise game options. This question was studied
 before for American options in \cite{CT} but the option price obtained there
 was not justified by classical hedging arguments. This justification was done
 in \cite{DIK} and \cite{IK1} for multiple exercise game options in the 
 discrete and continuous time cases, respectively, which by simplification 
 yields the result for American options, as well. This investigation required
 the study of Dynkin's games with multiple stopping which did not appear in the
 literature before.
 Observe that multiple exercise options may appear in their own rights
when an investor wants to buy or sell an underlying security in several 
instalments at times of his choosing and, actually, any usual American or
game option can be naturally extended to the multi exercise setup so that
they may emerge both in commodities, energy and in different financial markets.
Suppose, for instance, that a European car producer (having
most expenses in euros or pounds) plans to supply autos to US during a year in
several shipments and it buys a multiple exercise option which guaranties a 
favorable dollar--euro or pund exchange rate at time of shipments (of its 
choosing). The seller of such option can use currencies as underlying for his
hedging portfolio. A multiple exercise option could be cheaper then a basket 
of usual one exercise options if the former stipulates certain delay time 
between exercises which is quite natural in the above example. Furthermore,
the acting sides above may prefer to deal with game rather than American 
multiple exercise options since the former is cheaper for the buyer and safer
(because of cancellation clause) for the seller. 

Next, we describe more precisely game swing (multiple exercise) options in 
the CRR market where the stock price evolves according to (\ref{3.2}).
We consider a swing option of the game type which has the $i$-th payoff,
$i\geq 1$ having the form
\begin{equation}\label{3.9}
H^{(i)}(m,n)=X_i(m)\mathbb{I}_{m<n}+Y_i(n)\mathbb{I}_{n\leq{m}},
\,\,\,\forall{m,n}
\end{equation}
where $X_i(n),Y_i(n)$ are $\mathcal{F}_n$-adapted and
$0\leq Y_i(n)\leq X_i(n)<\infty$. Thus for any $i,n$ there
exist functions
$f^{(i)}_n,g^{(i)}_n:\{a,b\}^n\rightarrow{\mathbb{R}_{+}}$ such that
\begin{equation}\label{3.10}
Y_i(n)=f^{(i)}_n(\rho_1,...,\rho_n), \
X_i(n)=g^{(i)}_n(\rho_1,...,\rho_n).
\end{equation}
For any $1\leq{i}\leq{L-1}$ let $C_i$ be the set of all pairs
$((a_1,...,a_i),(d_1,...,d_i))\in {\{0,...,N\}}^i\times {\{0,1\}}^i$
such that $a_{j+1}\geq {N\wedge{(a_j+1)}}$ for any $j<i$. Such
sequences represent the history of payoffs up to the $i$-th one in
the following way. If $a_j=k$ and $d_j=1$ then the seller canceled
the $j$-th claim at the moment $k$ and if $d_j=0$ then the buyer
exercised the $j$-th claim at the moment $k$ (maybe together with
the seller). For $n\geq{1}$ denote by $\Gamma_n$ the set of all
stopping times with respect to the filtration
${\{\mathcal{F}_n\}}_{n=0}^N$ with values from $n$ to $N$ and set
$\Gamma=\Gamma_0$.

\begin{definition}\label{dfn3.1}
A stopping strategy is a sequence $s=(s_1,...,s_L)$  such that
$s_1\in\Gamma$ is a stopping time and for $i>1$,
$s_i:C_{i-1}\rightarrow{\Gamma}$ is a map which satisfies
$s_i((a_1,...,a_{i-1}),(d_1,...,d_{i-1}))\in
\Gamma_{N\wedge{(1+a_{i-1})}}$.
\end{definition}

In other words for the $i$-th payoff both the seller and the buyer
choose stopping times taking into account the history of payoffs so
far. Denote by $\mathcal{S}$ the set of all stopping strategies and
define the map
$F:\mathcal{S}\times\mathcal{S}\rightarrow{\Gamma^{L}\times\Gamma^{L}}$
by $F(s,b)=((\sigma_1,...,\sigma_L),(\tau_1,...,\tau_L))$ where
$\sigma_1=s_1$, $\tau_1=b_1$ and for $i>1$,
\begin{eqnarray*}
&\sigma_i=s_i((\sigma_1\wedge\tau_1,...,\sigma_{i-1}\wedge\tau_{i-1}),
(\mathbb{I}_{\sigma_1<\tau_1},...,
\mathbb{I}_{\sigma_{i-1}<\tau_{i-1}}))\,\,\,\mbox{and}\\
&\tau_i=b_i((\sigma_1\wedge\tau_1,...,\sigma_{i-1}\wedge\tau_{i-1}),
(\mathbb{I}_{\sigma_1<\tau_1},...,
\mathbb{I}_{\sigma_{i-1}<\tau_{i-1}})). \nonumber
\end{eqnarray*}
Set
\begin{equation*}
c_k(s,b)=\sum_{i=1}^L \mathbb{I}_{\sigma_i\wedge\tau_i\leq{k}}
\end{equation*}
which is a random variable equal to the number of payoffs until the
moment $k$.

For swing options the notion of a self financing portfolio involves not
only allocation of capital between stocks and the bank account but also
payoffs at exercise times. At the time $k$ the writer's decision how
much money to invest in stocks (while depositing the remaining money into
a bank account) depends not only on his present portfolio value but also
on the current claim. Denote by $\Xi$ the set of functions on the (finite)
probability space $\Omega$.
\begin{definition}\label{dfn3.2}
A portfolio strategy with an initial capital $x>0$ is a pair
$\pi=(x,\gamma)$ where
$\gamma:{\{0,...,N-1\}}\times{\{1,...,L\}}\times\mathbb{R}\rightarrow{\Xi}$
is a map such that $\gamma(k,i,y)$ is an $\mathcal{F}_k$-measurable
random variable which represents the number of stocks which the seller
buys at the moment $k$ provided that the current claim has the number $i$ and
the present portfolio value is $y$. At the same time the sum
$y-\gamma(k,i,y)S_k$ is deposited to the bank account of the portfolio. We
call a portfolio strategy $\pi=(x,\gamma)$ \textit{admissible} if for any
$y\geq{0}$,
\begin{equation}\label{3.11}
-\frac{y}{S_kb}\leq\gamma(k,i,y)\leq -\frac{y}{S_ka}.
\end{equation}
For any $y\geq{0}$ denote $K(y)=[-\frac{y}{b},-\frac{y}{a}]$.
\end{definition}

Notice that if the portfolio value at the moment $k$ is $y\geq{0}$
then the portfolio value at the moment $k+1$ before the payoffs (if
there are any payoffs at this time) is given by
$y+\gamma(k,i,y)S_k(\frac{S_{k+1}}{S_k}-1)$ where $i$ is the number
of the next payoff. In view of independency of
$\frac{S_{k+1}}{S_k}-1$ and $\gamma(k,i,y)S_k$ we conclude that the
inequality (\ref{3.11}) is equivalent to the inequality
$y+\gamma(k,i,y)S_k(\frac{S_{k+1}}{S_k}-1)\geq{0}$, i.e. the
portfolio value at the moment $k+1$ before the payoffs is nonnegative.
Denote by $\mathcal{A}(x)$ be the set of all \textit{admissible} portfolio
strategies with an initial capital $x>0$. Denote
$\mathcal{A}=\bigcup_{x>0}\mathcal{A}(x)$. Let $\pi=(x,\gamma)$ be a
portfolio strategy and $s,b\in\mathcal{S}$. Set
$((\sigma_1,...,\sigma_L),(\tau_1,...,\tau_L))=F(s,b)$ and
$c_k=c_k(s,b)$. The portfolio value at the moment $k$ after the
payoffs (if there are any payoffs at this moment) is given
by
\begin{eqnarray}\label{3.12}
&W^{(\pi,s,b)}_0=x-H^{(1)}(\sigma_1,\tau_1)
\mathbb{I}_{\sigma_1\wedge\tau_1=0} \ \ \mbox{and} \ \mbox{for} \
k>0, \\
&W^{(\pi,s,b)}_k=W^{(\pi,s,b)}_{k-1}+\mathbb{I}_{c_{k-1}<L}[\gamma(k-1,
c_{k-1}+1,W^{(\pi,s,b)}_{k-1})(S_k-S_{k-1})-
\nonumber\\
&\sum_{i=1}^L
H^{(i)}(\sigma_i,\tau_i)\mathbb{I}_{\sigma_i\wedge\tau_i=k}].\nonumber
\end{eqnarray}

\begin{definition}\label{dfn3.3}
A (perfect) hedge is a pair $(\pi,s)$ which consists of a portfolio
strategy and a stopping strategy such that $W^{(\pi,s,b)}_k\geq{0}$
for any $b\in\mathcal{S}$ and $k\leq{N}$.
\end{definition}
As usual, the option price $V^*$ is defined as the infimum of
$W\geq{0}$ such that there exists a hedge with an initial capital $W$.
The following result from \cite{DIK} provides a dynamical programming 
algorithm for computation of both the option price and the corresponding
hedge.

\begin{theorem}\label{thm3.2}
For any $n\leq{N}$ set
\begin{equation}\label{3.13}
X^{(1)}_n=X_L(n),\ Y^{(1)}_n=Y_L(n)\ and\  V^{(1)}_n=\min_{\sigma\in{\Gamma_{n}}}\max_{\tau\in{\Gamma_{n}}}
\tilde{E}(H^{(L)}(\sigma,\tau)|\mathcal{F}_n)
\end{equation}
and for $1<k\leq{L}$,
\begin{eqnarray}\label{3.14}
&X^{(k)}_n=X_{L-k+1}(n)+\tilde{E}(V^{(k-1)}_{(n+1)\wedge{N}}|\mathcal{F}_n),
\\
&Y^{(k)}_n=Y_{L-k+1}(n)+\tilde{E}(V^{(k-1)}_{(n+1)\wedge{N}}|\mathcal{F}_n)
\ \mbox{and} \nonumber\\
&V^{(k)}_n=\min_{\sigma\in{\Gamma_{n}}}\max_{\tau\in{\Gamma_{n}}}
\tilde{E}(X^{(k)}_{\sigma}\mathbb{I}_{\sigma<\tau}+Y^{(k)}_{\tau}
\mathbb{I}_{\sigma\geq\tau}
|\mathcal{F}_n)\nonumber
\end{eqnarray}
Where $\tilde E$ is the expectation with respect to the unique martingale
measure. Then
\begin{equation}\label{3.15}
V^{*}=V^{(L)}_0=\min_{s\in\mathcal{S}}\max_{b\in\mathcal{S}}G(s,b)
\end{equation}
where $G(s,b)=\tilde{E}\sum_{i=1}^L H^{(i)}(\sigma_i,\tau_i)$ and
$((\sigma_1,...,\sigma_L),(\tau_1,...,\tau_L))=F(s,b)$.
Furthermore, the stopping strategies
$s^{*}=(s^{*}_1,...,s^{*}_L)\in{S}$ and $b=(b^{*}_1,...,b^{*}_L)$
given by
\begin{eqnarray}\label{3.16}
&s^{*}_1=N\wedge{\min{\{k|X^{(L)}_k=V^{(L)}_k\}}}, \
b^{*}_1={\min{\{k|Y^{(L)}_k=V^{(L)}_k\}}},
\\
&s^{*}_i((a_1,...,a_{i-1}),(d_1,...,d_{i-1}))=N\wedge{\min{\{k>a_{i-1}|}}
\nonumber\\
&{{X^{(L-i+1)}_k=V^{(L-i+1)}_k\}}}, \
b^{*}_i((a_1,...,a_{i-1}),(d_1,...,d_{i-1}))\nonumber\\
&=N\wedge {\min{\{k>a_{i-1}| Y^{(L-i+1)}_k=V^{(L-i+1)}_k\}}}, \ i>1
\nonumber
\end{eqnarray}
satisfy the saddle point inequalities
\begin{equation}\label{3.17}
 G(s^{*},b)\leq G(s^{*},b^{*})\leq G(s,b^{*})\,\,\,
\mbox{for all}\,\, s,b
\end{equation}
and there exists a portfolio strategy
$\pi^{*}\in\mathcal{A}(V^{(L)}_0)$ such that $(\pi^{*},s^{*})$ is a hedge.
\end{theorem}

\section{Approximations of game options and of their shortfall risk}\label{sec4}

 Following \cite{Ki4} we will consider here approximations of the BS model
 by a sequence of CRR 
 models with the interest rates $r=r^{(n)}$ from (\ref{3.4})
 and with random variables $\rho_k=\rho_k^{(n)}$ from (\ref{3.2}) given by
 \begin{equation}\label{4.0}
 r=r^{(n)}=\exp(rT/n)-1\,\,\mbox{and}\,\,
 \rho_k=\rho_k^{(n)}=\exp\big(\frac {rT}{n}+\ka(\frac Tn)^{1/2}\xi_k\big)-1
 \end{equation} 
 where $\xi_j=\xi_j^{(n)},\, j=1,2,...$ are i.i.d. random variables taking on
 the values $1$ and $-1$ with probabilities $p^{(n)}=(\exp(\ka\sqrt{\frac Tn})
 +1)^{-1}$ and $1-p^{(n)}=(\exp(-\ka\sqrt{\frac Tn})+1)^{-1}$, respectively. 
 This choice of random variables $\xi_i,\, i\in\bbN$ determines already the 
 probability measures $P^\xi_n=\{ p^{(n)},1-p^{(n)}\}^\infty$ for the above 
 sequence of CRR models and since $E_n^\xi\rho_k^{(n)}=r^{(n)}$, where
 $E_n^\xi$ is the expectation with respect to $P^\xi_n$, we conclude that
 $P^\xi_n$ is the martingale measure for the corresponding CRR market and
 the fair price $V=V^{(n)}$ of a game option in this market is given by the
 formula (\ref{3.5}) with $E=E_n^\xi$.
  
 Let $V$ be the fair price of the game option
 in the BS market. It turns out that for a certain
 natural class of payoffs $X_t$ and $Y_t$ which may depend on the whole path
 (history) of the stock price evolution (as in integral or Russian type 
 options) the error $|V-V^{(n)}|$ does not exceed $Cn^{-1/4}(\ln n)^{3/4}$
 where $C>0$ does not depend on $n$ and it can be estimated
  explicitly. Moreover, we will see that the rational exercise times of 
  our CRR binomial approximations yield near rational 
  ($Cn^{-1/4}(\ln n)^{3/4}$-optimal stopping times for the corresponding
  Dynkin games) exercise times for 
  game options in the BS market. Since the values $V^{(n)}$ and the optimal
  stopping times of the corresponding discrete time Dynkin's games can be
  obtained directly via the dynamical programming recursive procedure
  (\ref{3.7}) our results provide a justification of a rather effective
   method of computation of fair prices and exercise times of game options
   with path dependent payoffs. The standard construction of a 
   self-financing hedging 
   portfolio involves usually the Doob--Meyer decomposition of supermartingales
   which is explicit only in the discrete but not in the continuous time case.
   We will see how to construct a self-financing portfolio in the BS market 
   with a small average (maximal) shortfall and an initial capital close to
   the fair price of a game option using hedging self-financing portfolios
   for the approximating binomial CRR markets. The latter problem does not 
   seem to have been addressed before \cite{Ki4} in the literature on this 
   subject.
   This hints, in particular, that since hedging self-financing portfolio
   strategies can be computed only approximately their possible shortfalls
   come naturally into the picture and they should be taken into account in
   option pricing even if a perfect hedging exists theoretically. Note that
    these results require not only
   an approximation of stock prices and the corresponding payoffs but also 
   we have to take care about the different nature of stopping times in 
   (\ref{3.5}) and (\ref{3.6}). 
  
  The main tool here is the Skorokhod type embedding (see \cite{Bil}) of sums 
  of i.i.d. random variables into a Brownian motion (with a constant drift, 
  in this case).
  This tool was already employed for similar purposes in \cite{LR} and in 
  \cite{Wa}. The first paper treats an optimal stopping problem which can 
  be applied to an American style option
   with a payoff function depending only on the current stock price and,
   more importantly, this function must be bounded and have two bounded 
   derivatives which excludes usual put and call options cases. The second 
   paper deals only with European options and, again, only payoffs (though with
   some discontinuities) determined by the current stock price are allowed.
   We observe that the Skorokhod embedding does not provide optimal error
   estimates in strong approximation theorems and it would be interesting
   to understand whether other approaches such as the quantile method
   (see \cite{Za}) and Stein's method (see \cite{Ch}) can be employed 
   for approximation of optimal stopping game values with better estimates
   of errors. Skorokhod embedding does not work also in the multidimensional
   situation and for this case another method from \cite{DP} was employed in
   \cite{Ki5} where, actually, more general and not only binomial 
   approximations were considered.
   More general approximation results for game options were obtained in 
   \cite{Do1} where only continuity of payoffs were assumed but as a result
   no error estimates could be obtained there.

For each $t>0$ denote by $M[0,t]$  the space of Borel measurable functions 
on $[0,t]$  with the uniform metric $d_{0t}(\up,\tilde\up)=
\sup_{0\leq s\leq t}|\up_s-\tilde\up_s|$. For each $t>0$ let $F_t$ and 
$\Del_t$ be nonnegative functions on $M[0,t]$ such that for some constant
$L\geq 1$ and for any $t\geq s\geq 0$ and $\up,\tilde\up\in M[0,t]$,
\begin{equation}\label{4.1}
|F_s(\up)-F_s(\tilde\up)|+|\Del_s(\up)-\Del_s(\tilde\up)|\leq L(s+1)d_{0s}
(\up,\tilde\up),
\end{equation}
and
\begin{equation}\label{4.2}
|F_t(\up)-F_s(\up)|+|\Del_t(\up)-\Del_s(\up)|\leq L\big(|t-s|
(1+\sup_{u\in[0,t]}|\up_u|)+\sup_{u\in[s,t]}|\up_u-\up_s|\big).
\end{equation}
By (\ref{4.1}), $F_0(\up)=F_0(\up_0)$ and $\Del_0(\up)=\Del_0(\up_0)$
are functions of $\up_0$ only. By (\ref{4.2}),
\begin{equation}\label{4.3}
F_t(\up)+\Del_t(\up)\leq F_0(\up_0)+\Del_0(\up_0)+L(t+2)
(1+\sup_{0\leq s\leq t}|\up_s|)
\end{equation}

Next, we consider the BS market on a complete probability space together
with its martingale measure $P^B$ which exists and is unique as a corollary
of the Girsanov theorem (see \cite{Sh}).
Let $B_t,\, t\geq 0$ be the standard one-dimensional continuous in time 
Brownian motion with respect to the martingale measure $P^B$. Set
\[
B_t^*=-\frac \ka 2t+B_t,\quad t\geq 0.
\]
Then the stock price $S^B_t(z)$ at time $t$ in the BS market can be written 
in the form
\begin{equation}\label{4.4}
S_t^B(z)=z\exp(rt+\ka B^*_t),\,\,\, S^B_0(z)=z>0,
\end{equation}
where $r>0$ is the interest rate and $\ka>0$ is the, so called, volatility.
We will consider game options in the BS market with payoff processes in the 
form
\begin{equation*}
Y_t=F_t(S^B(z))\quad\mbox{and}\quad X_t=G_t(S^B(z)),\,\, t\in[0,T],\, T>0,
\end{equation*}
where $G_t=F_t+\Del_t$, $F,\Del$ satisfy (\ref{4.1}) and (\ref{4.2}), 
$S^B(z)=S^B(z,\om)\in M[0,T]$ is a random function taking the value 
$S_t^B(z)=S_t^B(z,\om)$ at $t\in[0,T]$, and in the notations $F_t(S^B(z))$
$G_t(S^B(z))$ for $t<T$ we take the restriction of $S^B(z)$ to the interval
$[0,t]$. The fair price $V=V(z)$ of this option with an initial value $z>0$
of the stock is given by (\ref{3.6}).

Next, we consider a sequence of CRR markets on a complete probability
space such that for each $n=1,2,...$
the stock prices $S_t^{(n)}(z)$ at time $t$ are given by the formula
\begin{eqnarray}\label{4.5}
&S_t^{(n)}(z)=z\exp\bigg(\sum_{k=1}^{[nt/T]}\big(\frac {rT}{n}+\ka(\frac
Tn)^{1/2}\xi_k\big)\bigg),\,\,\, t\geq T/n\\
&\mbox{and}\,\,\, S_t^{(n)}(z)=S^{(n)}_0(z)=z>0,\,\,\,\, t\in[0,T/n)\nonumber
\end{eqnarray}
where, recall, $\xi_1,\,\xi_2,...$ are i.i.d. random variables taking the 
values 1 and $-1$ with probabilities $p^{(n)}=(\exp(\ka\sqrt{\frac Tn})
 +1)^{-1}$ and $1-p^{(n)}=(\exp(-\ka\sqrt{\frac Tn})+1)^{-1}$, respectively. 
 Namely, we consider CRR markets where stock prices $S_m=S^{(n)}_{\frac mn}(z),
 \, m=0,1,2,...$ satisfy (\ref{3.2}) with $\rho_k=\rho_k^{n}$ given by 
 (\ref{4.0}) and, in addition, in place of the interest rate $r$ in the first 
formula in (\ref{3.4}) we take the sequence of interest rates $r_n=\exp(rT/n)-1$
where $r$ is the interest rate of the BS market appearing in the second 
formula of (\ref{3.4}) and in (\ref{3.6}). We consider $S^{(n)}(z)=
S^{(n)}(z,\om)$ as a random function on $[0,T]$, so that $S^{(n)}(z,\om)\in
M[0,T]$ takes the value $S^{(n)}_t(z)=S^{(n)}_t(z,\om)$ at $t\in[0,T]$.
For $k=0,1,2,...,n$ put
\begin{equation}\label{4.6}
Y_k=Y_k^{(n)}(z)=F_{\frac {kT}{n}}(S^{(n)}(z))\,\,\,\mbox{and}\,\,\,
X_k=X_k^{(n)}(z)=G_{\frac {kT}{n}}(S^{(n)}(z)).
\end{equation}
Then for each $n$ the fair price $V=V^{(n)}(z)$ of the game option in the 
corresponding CRR market with an initial value $z>0$ of the stock is given 
by (\ref{3.5}). 

Set
\begin{equation}\label{4.8}
H_z^B(s,t)=F_t(S^B(z))\bbI_{s\geq t}+G_s(S^B(z))\bbI_{s<t}\,\,\mbox{and},\,\,
Q_z^B(s,t)=e^{-rs\wedge t}R_z^B(s,t)\,\,\mbox{and}\,\,
\end{equation}
\begin{equation}\label{4.9}
H_z^{(n)}(s,t)=F_t(S^{(n)}(z))\bbI_{s\geq t}+G_s(S^{(n)}(z))\bbI_{s<t},
\,\, Q_z^{(n)}(s,t)=e^{-rs\wedge t}R_z^{(n)}(s,t).
\end{equation}
 Denote by $\cT^B_{0T}$ and $\cT_{0n}^\xi$ the sets of stopping times with 
respect to the Brownian filtration $\cF_t^B,\, t\geq 0$ with values in 
$[0,T]$ and with respect to the filtration $\cF_k^\xi=\sig\{\xi_1,...,\xi_k\}$
 with values in $\{ 0,1,...,n\}$. Set
\begin{equation}\label{4.10}
V(z)=\inf_{\sig\in\cT^B_{0T}}\sup_{\tau\in\cT^B_{0T}}E^BQ_z^B(\sig,\tau)\,\,
\mbox{and}\,\,
\end{equation}
\begin{equation}\label{4.11}
V^{(n)}(z)=\min_{\zeta\in\cT_{0n}^\xi}\max_{\eta\in\cT_{0n}^\xi}
E^\xi_nQ_z^{(n)}(\frac {\zeta T}n,\frac {\eta T}n).
\end{equation}
where $E^B$ and $E^\xi_n$ are the expectations with respect to
the probability measures $P^B$ and $P^\xi_n$, respectively,
and we observe that $\cT_{0n}^\xi$ is finite set
 so that we can use $\min$ and $\max$ in (\ref{4.11}).
  
 Recall, that we choose $P^B$ to be the martingale measure for the BS market
 and observe that $P^\xi_n$ is the martingale measure for the corresponding
 CRR market since a direct computation shows that $E^\xi_n\rho_k^{(n)}=r_n$.
 Thus, (\ref{4.10}) and (\ref{4.11}) give fair prices of the game options in
 the corresponding markets. 
 We note also that all our formulas involving the expectations $E^B$, in 
 particular, (\ref{4.10}) giving the fair price $V$ of a game option, do not
 depend on a particular choice of a continuous in time version of the Brownian
 motion since all of them induce the same probability measure on the space
 of continuous sample paths which
 already determines all expressions with the expectations $E^B$ appearing
 in this paper.
 
The following result from \cite{DK2} 
provides an estimate for the error term in approximation
of the fair price of a game option in the BS market by fair prices of the
sequence of game options and prices of Dynkin's games defined above.

\begin{theorem}\label{thm4.1} Suppose that $V(z)$ and $V^{(n)}(z)$ are defined
by (\ref{4.9})--(\ref{4.11}) with functions $F$ and $G=F+\Del$ satisfying
(\ref{4.1}) and (\ref{4.2}). Then there exists a constant $C>0$
(which can be explicitly estimated) such that
\begin{equation}\label{4.13}
|V(z)-V^{(n)}(z)|\leq C (F_0(z)+\Del_0(z)+z+1)n^{-\frac 14}(\ln n)^{3/4}
\end{equation}
for all $z,n>0$.
\end{theorem}

 We can choose more general i.i.d.
random variables $\xi_1,\xi_2,...$ appearing in the definition of $V^{(n)}$,
as well, but these generalizations do not seem to have a
financial mathematics motivation since we want to approximate game options
in the BS market by simplest possible models which are, of course, game
options in the CRR market.

Among main examples of options with path-dependent payoff we have in mind
integral options where
\[
F_t(\up)=(\int_0^tf_u(\up_u)du-L)^+\quad\mbox{(call option case)}
\]
or
\[
F_t(\up)=(L-\int_0^tf_u(\up_u)du)^+\quad\mbox{(put option case)}
\]
where, as usual, $a^+=\max(a,0)$. The penalty functional may also have
here the integral form
\[
\Del_t(\up)=\int_0^t\del_u(\up_u)du.
\]
In order to satisfy the conditions (\ref{4.1}) and (\ref{4.2}) we can assume
that for some $K>0$ and all $x,y,u$,
\[
|f_u(x)-f_u(y)|+|\del_u(x)-\del_u(y)|\leq K|x-y|
\]
and
\[
|f_u(x)|+|\del_u(x)|\leq K|x|.
\]

Observe, also that the Asian type (averaged integral) payoffs of the form
\[
F_t(\up)=(\frac 1t\int_0^tf_u(\up_u)du-L)^+\,\,\mbox{or}\,\, 
=(L-\frac 1t\int_0^tf_u(\up_u)du)^+
\]
do not satisfy the condition (\ref{4.2}) if arbitrarily small exercise times
are allowed though the latter seems to have only some theoretical interest as
it hardly happens in reality. Still, also in this case the binomial 
approximation errors can
be estimated in a similar way considering separately estimates for small 
stopping times and for stopping times bounded away from zero. Namely,
define $V_\ve(z)$ and $V_\ve^{(n)}(z)$ for $\ve\geq 0$
by (\ref{4.10}) and (\ref{4.11})
where $Q^{(B)}_z(\sig,\tau)$ and $Q^{(n)}_z(\frac {\zeta T}n,\frac {\eta T}n)$
are replaced by $Q^{(B)}_z(\sig\vee\ve,\tau\vee\ve)$ and
 $Q^{(n)}_z(\frac {\zeta T}n\vee\ve,\frac {\eta T}n\vee\ve)$, respectively.
 Assuming that $f_u$ and $\del_u$ are Lipschitz continuous also in $u$
 (at least for $u$ close to 0) in the form $|f_s(x)-f_u(x)|+|\del_s(x)-
 \del_u(x)|\leq K(x+1)|s-u|$ for some $K>0$ and all $s,u,x\geq 0$ we obtain 
 that if $\up_0=z$ and $F_0(\up)=(f_0(z)-L)^+$ or $=(L-f_0(z))^+$ then
 \[
 |F_s(\up)-F_0(z)|\leq Ks(1+\sup_{0\leq u\leq s}|\up_u|)+
 K\sup_{0\leq u\leq s}|\up_u-z|.
 \]
 It is not difficult to
 see from here that $|V(z)-V_\ve(z)|$ and $|V^{(n)}(z)-V_\ve^{(n)}(z)|$ do
 not exceed $C(1+z)\sqrt\ve$ for all small $\ve$ and some constant $C$. On the
 other hand, similarly to Theorem 
 \ref{thm4.1} we see that for some constant $C>0$ and all $n,\ve>0$,
 \[
 |V_\ve(z)-V_\ve^{(n)}(z)|\leq C(1+z)\ve^{-1}n^{-1/4}
 (\ln n)^{3/4}.
 \]
 Choosing $\ve=n^{-1/6}\sqrt {\ln n}$ we obtain that under the above conditions
 in the case of Asian options $|V(z)-V^{(n)}(z)|$ can be estimated by
  $3C(1+z)n^{-1/12}(\ln n)^{1/4}$.

Another important example of path-dependent payoffs are, so called, Russian
options where, for instance,
\[
F_t(\up)=\max\big(m,\sup_{u\in[0,t]}\up_u\big)\,\,\mbox{and}\,\,
\Del_t(\up)=\del\up_t.
\]
Such payoffs satisfy the conditions of Theorem \ref{thm4.1}. Indeed, (\ref{4.1})
is clear in this case and (\ref{4.2}) follows since for $t\geq s$,
\begin{eqnarray*}
&\max(m,\sup_{u\in[0,t]}v_u)-\max(m,\sup_{u\in[0,s]}v_u)\leq
\sup_{u\in[0,t]}v_u-\sup_{u\in[0,s]}v_u\\
&\leq\sup_{u\in[s,t]}v_u-v_s\leq\sup_{u\in[s,t]}|v_u-v_s|.
\end{eqnarray*}

In order to compare $V(z)$ and $V^{(n)}(z)$ in the case of path dependent 
payoffs we have to consider both BS and CRR markets on one probability space 
in an appropriate way and the main tool in achieving
this goal will be here the Skorokhod type embedding (see, for instance, 
\cite{Bil}, Section 37). In fact, for the binomial i.i.d. random variables 
$\xi_1,\xi_2,...$ appearing in the setup of
the CRR market models above the embedding is explicit and no general theorems
are required but if we want to extend the result for other sequences of i.i.d.
random variables we have to rely upon the general result. Namely,
define recursively
\[
\te_0^{(n)}=0,\,\te_{k+1}^{(n)}=\inf\{ t>\te_k^{(n)}:\, 
|B^*_t-B^*_{\te_k^{(n)}}|=\sqrt {\frac Tn}\},
\]
where, recall, $B^*_t=-\frac \ka 2t+B_t$.
The standard strong Markov property based arguments (cf. \cite{Bil}, 
Section 37) show that $\te_k^{(n)}-\te_{k-1}^{(n)},\, k=1,2,...$  are i.i.d.
 sequences of random variables such that $(\te^{(n)}_{k+1}-\te_k^{(n)},\,
B^*_{\te_{k+1}^{(n)}}-B^*_{\te_k^{(n)}})$  are
independent of $\cF^B_{\te_k^{(n)}}$ (where, recall, $\cF^B_t=\sig\{ B_s,
\, s\leq t\}$).

 It turns out (see \cite{Wa} and \cite{Ki4}) that
$B^*_{\te_1^{(n)}}$ has the same distribution as $\sqrt {\frac Tn}\xi_1$. Set
\begin{equation}\label{4.15}
\Xi_k^{(n)}=(\frac Tn)^{1/2}\sum_{j=1}^k\xi_j,
\end{equation}
then $\Xi_k^{(n)}$ has the same distribution as $B^*_{\te_k^{(n)}}$.

Theorem \ref{thm4.1} provides an approximation of the fair price of game
options in the BS market by means of fair prices of game options in the 
CRR market which becomes especially useful if we can provide also a simple
description of rational (or $\del$-rational) excercise times of these options 
in the BS market via exercise times of their CRR market approximations
which are, by the definition, optimal (or $\del$-optimal) stopping times
for the Dynkin game whose price is given by (\ref{4.11}).
For each $k=1,2,...$ introduce the finite $\sig$-algebra $\cG_k^{B,n}=
\sig\{ B^*_{\te_1^{(n)}},B^*_{\te_2^{(n)}}-B^*_{\te_1^{(n)}},...,
B^*_{\te_k^{(n)}}-B^*_{\te_{k-1}^{(n)}}\}$ which is, clearly, isomorphic to 
$\cF_k^\xi=\sig\{\xi_i,\, i\leq k\}$ considered before since each element
of $\cG_k^{B,n}$ and of $\cF_k^\xi$ is an event of the form
\[ 
A^{B,n}_{\iota^{(k)}}=\{ B^*_{\te_j^{(n)}}-B^*_{\te_{j-1}^{(n)}}=
\iota_j\sqrt\frac Tn,\, j=1,...,k\}
\]
 and 
\[
 A^\xi_{\iota^{(k)}}=\{\xi_j=\iota_j,\, j=1,...k\},
 \]
  respectively, where $\iota^{(k)}=(\iota_1,...,\iota_k)\in\{ -1,1\}^k$, 
  $\te^{(n)}_0 =0$ and $B_0=0$. Let $\cS^{B,n}$ be the set of stopping times 
  with respect to the filtration $\cG_k^{B,n},\, k=0,1,2,...$, where 
  $\cG_0^{B,n}= \{\emptyset,\Om_B\}$ is the trivial $\sig$-algebra and
  $\Om_B$ is the sample space of the Brownian motion. The subset of these
  stopping times with values in $\{0,1,...,n\}$ will be denoted by 
  $\cS^{B,n}_{0,n}$. For each 
  $\iota^{(n)}=(\iota_1,...,\iota_n)\in\{ -1,1\}^n$ and $k<n$ we set 
  $\iota^{(k)}=
  (\iota_1,...,\iota_k)\in\{ -1,1\}^k$. Denote by $\cJ_{0,n}$ the set of 
  functions
  $\nu :\{ -1,1\}^n\to\{ 0,1,...,n\}$ such that if $\nu(\iota^{(n)})=k\leq n$
  and $\tilde\iota^{(k)}=\iota^{(k)}$ for some $\tilde\iota^{(n)}\in\{ -1,1\}^n$
   then $\nu(\tilde\iota^{(n)})=k$, as well. Define the functions
  $\la^{(n)}_\xi:\Om_\xi\to\{ -1,1\}^n$ and $\la^{(n)}_B:\Om_B\to\{ -1,1\}^n$ by
  $\la^{(n)}_\xi(\om)=(\xi_1(\om),...,\xi_n(\om))$ and
  \[
  \la^{(n)}_B(\om)=\sqrt\frac nT\big(B^*_{\te_1^{(n)}(\om)}(\om),
  B^*_{\te_2^{(n)}(\om)}(\om)-B^*_{\te_1^{(n)}(\om)}(\om),...,
  B^*_{\te_n^{(n)}(\om)}(\om)-B^*_{\te_{n-1}^{(n)}(\om)}(\om)\big)
  \]
  where $\Om_\xi$ and $\Om_B$ are sample spaces on which the sequence 
  $\xi_1,\xi_2,...$ and the Brownian motion $B_t$ are defined, respectively.
  It is clear that any $\zeta\in\cT^\xi_{0n}$ and $\eta\in\cS^{B,n}_{0,n}$ can
  be represented uniquely in the form $\zeta=\mu\circ\la^{(n)}_\xi$ and 
  $\eta=\nu\circ\la^{(n)}_B$ for some $\mu,\nu\in\cJ_{0,n}$. 
   
  \begin{theorem}\label{thm4.2}
  There exists a constant $C>0$ (which can be estimated
  explicitly) such that if $\zeta_n^*=\mu^*_n\circ\la^{(n)}_\xi$ 
  and $\eta^*_n=\nu^*_n\circ\la^{(n)}_\xi$, $\mu^*_n,\nu^*_n\in\cJ_{0n}$ 
  are rational exercise times for the game option in the CRR market defined by
  (\ref{4.5}), i.e.
  \begin{equation}\label{4.16}
  V^{(n)}(z)=\min_{\zeta\in\cT_{0n}^\xi}E^\xi Q_z^{(n)}(\zeta\frac Tn,
\eta^*_n\frac Tn)=\max_{\eta\in\cT_{0n}^\xi}
E^\xi Q_z^{(n)}(\zeta^*_n\frac Tn,\eta\frac Tn)
\end{equation}
then $\vf^*_n=\te^{(n)}_{\mu^*_n\circ\la^{(n)}_B}$ and 
$\psi^*_n=\te^{(n)}_{\nu^*_n\circ\la^{(n)}_B}$ are $\del_n(z)$-rational
exercise times for the game option in the BS market defined by (\ref{4.3}) and 
(\ref{4.4}), i.e.
\begin{equation}\label{4.17}
\sup_{\tau\in\cT^B_{0T}}E^BQ_z^B(\vf^*_n,\tau)-\del_n(z)\leq V(z)
 \leq\inf_{\sig\in\cT^B_{0T}}E^BQ_z^B(\sig,\psi^*_n)+\del_n(z),
\end{equation}
where $\del_n(z)=C (F_0(z)+\Del_0(z)+z+1)n^{-\frac 14}(\ln n)^{3/4}$. 
\end{theorem}

It is well known (see, for instance, \cite{Ne}) that when payoffs depend 
only on the current stock price (a Markov case) $\del$-optimal stopping times
of Dynkin's games can be obtained as first arrival times to domains where
the payoff is $\del$-close to the value of the game (as a function of the
initial stock price). For path dependent payoffs the situation
is more complicated and, in general, in order to construct $\del$-optimal 
stopping times we have to know the stochastic process of values of the games
 starting at each time $t\in[0,T]$ conditioned to the information up to $t$. 
 It is not clear what kind of approximation of this process can provide some
 information about $\del$-rational exercise times and the convenient 
 alternative method of their construction exhibited in Theorem \ref{thm4.2}
 seems to be important both for the theory and applications. Moreover,
 this construction is effective and can be employed in practice since  
$\mu^*_n$ and $\nu^*_n$ are functions on sequences of $1$'s and $-1$'s
which can be computed (and stored in a computer) using the recursive 
formulas (\ref{3.7}) even before the stock evolution begins. In order 
to compute $\la_B^{(n)}$ we have to watch the discounted stock price 
$\check S^B_t(z)=e^{-rt}S^B_t(z)$ evolution of a real stock at moments 
$\te^{(n)}_k$ which are obtained recursively by $\te^{(n)}_0=0$ and
 \begin{equation}\label{4.18}
 \te^{(n)}_{k+1}=\inf\{ t>\te^{(n)}_k:\,\check S^B_t(z)=
 e^{\pm\ka(\frac Tn)^{1/2}}\check S^B_{\te^{(n)}_k}(z)\}
 \end{equation} 
 and to construct the $\{ 1,-1\}$ sequence $\la_B^{(n)}(\om)$ by writing
 $1$ or $-1$ on $k$th place depending on whether
  $\check S^B_{\te^{(n)}_k}(z)=e^{\ka(\frac Tn)^{1/2}}
  \check S^B_{\te^{(n)}_{k-1}}(z)$ or $\check S^B_{\te^{(n)}_k}(z)=
  e^{-\ka(\frac Tn)^{1/2}}\check S^B_{\te^{(n)}_{k-1}}(z)$,
 respectively. 
 
 Recall (see \cite{Sh}), that a sequence $\pi=(\pi_1,...,\pi_n)$ of pairs
  $\pi_k=(\be_k,\gam_k)$ of $\cF^\xi_{k-1}$-measurable random variables
  $\be_k,\,\gam_k,\, k=1,...,n$ is called a self-financing portfolio
  strategy in the CRR market determined by (\ref{3.2}), (\ref{3.4}),
  (\ref{4.0}) and (\ref{4.5}) if the price of the portfolio at time $k$
  is given by the formula
  \begin{equation}\label{4.19}
  W_k^{\pi,n}=\be_kb_k+\gam_kS^{(n)}_{\frac {kT}n}(z)=\be_{k+1}b_k+
  \gam_{k+1}S^{(n)}_{\frac {kT}n}(z)
  \end{equation}
  and the latter equality means that all changes in the portfolio value 
  are due to capital gains and losses
  but not to withdrawal or infusion of funds. A pair $(\zeta,\pi)$ of a
  stopping time $\zeta\in\cT^\xi_{0n}$ and a self-financing portfolio strategy
  $\pi$ is called a hedge for (against) the game option with the payoff 
  $R^{(n)}_z$ given by (\ref{4.9}) if (see \cite{Ki3}),
  \begin{equation}\label{4.20}
  W_{\zeta\wedge k}^{\pi,n}\geq H^{(n)}_z(\frac {\zeta T}n,\,\frac {kT}n),
  \,\,\,\forall 
  k=0,1,...,n.
  \end{equation}
  It follows from \cite{Ki3} that for any $\zeta\in\cT^\xi_{0n}$ there exists
  a self-financing portfolio strategy $\pi^\zeta$ so that $(\zeta,\pi^\zeta)$
  is a hedge. In particular, if we take the rational exercise time 
  $\zeta=\zeta^*_n$ of the writer then such $\pi^\zeta$ exists with the
  initial portfolio capital $V^{(n)}(z)$. The construction of $\pi^{\zeta}$ 
  goes directly via the Doob decomposition of supermartingales and a martingale
  representation lemma (see \cite{Sh} and \cite{Ki3}) both being explicit in
  the CRR market case. In the continuous time BS market we cannot write the
  corresponding portfolio strategies in an explicit way, and so some
  approximations are necessary.

  \begin{theorem}\label{thm4.3}
  Let $\zeta\in\cT^\xi_{0n}$, $\pi=\pi^\zeta$ and (\ref{4.19}) together
  with (\ref{4.20}) hold true with $\cF_k^\xi$-measurable $\be_k=\be_k^\zeta$
   and $\gam_k=\gam_k^\xi$, so that $(\zeta,\pi^\zeta)$ is a hedge. Then
  $\be_k^\zeta=f_k\circ\la_\xi^{(k-1)}$, $\gam_k^\zeta=g_k\circ\la_\xi^{(k-1)}$,
  and $\zeta=\mu\circ\la_\xi^{(n)}$ for some uniquely defined functions $f_k$,
  $g_k$ on $\{-1,1\}^{k-1}$ and some $\mu\in\cJ_{0n}$. Let $\vf=\mu\circ\la_B^{(n)}$ and
  set $\be^\vf_t=f_k\circ\la_B^{(k-1)}$ and $\gam_t^\vf=g_k\circ\la_B^{(k-1)}$
  whenever $t\in(\te_{k-1}^{(n)},\te_{k}^{(n)}]$. Then
  \begin{equation}\label{4.21}
  W_t^B=\be_t^\vf b_t+\gam_t^\vf S_t^B(z)
  \end{equation}
  is a self-financing portfolio in the BS market and there exists a constant
  $C>0$ such that
  \begin{equation}\label{4.22}
  E^B\sup_{0\leq t\leq T}\big(H^B_z(\te_\vf^{(n)},t)-
  W^B_{\te_\vf^{(n)}\wedge t}\big)^+\leq C(F_0(z)+\Del_0(z)+z+1)
  n^{-\frac 14}(\ln n)^{3/4}
  \end{equation}
  where $a^+=\max(a,0)$. In particular, there exists a self-financing portfolio
  of this form satisfying (\ref{4.22}) with the initial value $V^{(n)}(z)$
  (which according to (\ref{4.13}) is close to the fair price
   $V(z)$ of the game option) taking $\vf^*=\mu^*\circ\la_B^{(n)}$ if
   $\zeta^*=\mu^*\circ\la_\xi^{(n)}$ is the rational exercise time and
   $\pi=\pi^{\zeta^*}$ is the corresponding optimal self-financing hedging
   portfolio strategy for the CRR market.
  \end{theorem}
  
  The inequality (\ref{4.22}) estimates the expectation of the maximal
  shortfall (risk) of certain (nearly hedging) portfolio strategy which can be 
  constructed effectively in applications since the functions $f_l,g_l$, and
  $\mu$ are determined by a self-financing hedging strategy in the CRR market
  which can be computed directly and stored in a computer even before the real
  stock evolution begins or in case of computer memory limitations we can 
  compute these functions each time when needed using corresponding
  algorithms for the CRR market. The functions $\la_B^{(n)}$ or, in other
  words, the sequences from $\{-1,1\}^n$ which should be pluged into the 
  functions $f_l$, $g_l$, and $\mu$ should be obtained in practice by 
  watching the evolution of the discounted stock price $e^{-rt}S_t^B$ 
  at moments $\te^{(n)}_k$ as described after Theorem \ref{thm4.2}.
  
  The paper \cite{DK2} studied approximations of the shortfall risk $R(x)$
   given by (\ref{3.8}) for game options in the BS market by the shortfall
   risks $R_n(x)$ of game options in the sequence of CRR markets defined
   above where the initial capital $x$ of all portfolios under consideration
   is kept the same and the payoffs satisfy the same conditions as above.
   The convergence $\lim_{n\to\infty}R_n(x)=R(x)$ was proved in \cite{DK2}
   but only the one sided error estimate 
   \[
   R(x)\leq R_n(x)+Cn^{-1/4}(\ln n)^{3/4}
   \]
   was obtained there for game options while relying on some convexity
   arguments it was possible to obtain for American options two sided
   estimates with the same error term. 
   
 In \cite{DK3} similar approximation results as above were extended to
 barrier game options. Namely, \cite{DK3} deals with double knock--out 
 barrier option with two constant barriers $L,R$ such that 
 $0\leq L<S_0<R\leq\infty$ which means that the option becomes
worthless to its holder (buyer) at the first time $\tau_I$
the stock price $S_t$ exits the open interval $I=(L,R)$. Thus for
$t\geq\tau_{(L,R)}$ the payoff is $X_t=Y_t=0$. For $t<\tau_{(L,R)}$
path dependent payoffs satisfying (\ref{4.2}) and (\ref{4.3}) are considered.
Such a contract is of potential value to a buyer who believes that the stock
price will not exit the interval $I$ up to a maturity date and to a seller 
who does not want to worry about hedging if the stock price will reach one
of the barriers $L,R$. Such an option is equivalent to the usual game option
when the payoffs $X_t$ and $Y_t$ are replaced by $X_t^I=X_t\bbI_{t<\tau_I}$
and $Y_t^I=Y_t\bbI_{t<\tau_I}$, respectively. Now, these new payoffs loose
regularity conditions (\ref{4.2}) and (\ref{4.3}) but still it turns out that
the error estimates in (\ref{4.13}) remain true when we approximate the price
of the above barrier game options in the BS market by the prices of 
corresponding barrier game options in the CRR markets as in Theorem 
\ref{thm4.1} above. The results concerning approximation of the shortfall
risk turn out to be very similar for barrier game options as the corresponding
results for usual game options described above.

When payoffs depend only on the current stock price (and not path dependent 
as in (\ref{4.8}) and (\ref{4.9})) then in some special cases it is possible
to obtain
better then in Theorem \ref{thm4.1} error estimates for binomial approximations
of prices of game options relying on partial differential equations methods
in the free boundary problem. In \cite{La} this was done for American put
options in the BS market and in \cite{IK2} this was extended to game put 
options with error estimates of order $n^{-1/2}$ in comparison to 
$n^{-1/4}(\ln n)^{3/4}$ obtained in Theorem \ref{thm4.1}.

  \section{Incomplete markets and transaction costs}\label{sec5}

Both in incomplete markets and in markets with transaction costs there
is no one arbitrage free price of each derivative which can be considered
as its fair price and one of approaches in these circomstances is to study
superhedging. Game options in incomplete markets were studied in several
papers, in particular, in \cite{Ku} they were studied from the point of
view of utility maximization which leads to non-zero-sum Dynkin's games
while in \cite{KaK1}) they were studied from the point of view of superhedging
and arbitrage free prices.

Next, we concentrate in this section in the superhedging pricing of game
options in discrete markets with transactions costs. The market model here
will consists of a finite probability space $\Om$ with the $\sig$-field 
$\cF=2^\Om$ of all subspaces of $\Om$ and a probability measure $\bbP$ on 
$\cF$ giving a positive weight $\bbP(\om)$
  to each $\om\in\Om$. The setup includes also a filtration $\{\emptyset,
  \Om\}=\cF_0\subset\cF_1\subset ...\subset\cF_T=\cF$ where $T$ is a positive
   integer called the time horizon. It is convenient to denote by $\Om_t$ the
   set of atoms in $\cF_t$ so that any $\cF_t$-measurable random variable 
   (vector) $Z$ can be identified with a function (vector function) defined on
    $\Om_t$ and its value at $\mu\in\Om_t$ will be denoted either by $Z(\mu)$
    or by $Z^\mu$.
    
     The market model consists of a risk-free bond and a risky stock. 
     Without loss of generality, we can assume that all prices are discounted
     so that the bond price equals 1 all the time and a position in bonds is
     identified with cash holding. On the other hand, the shares of the stock
     can be traded which involves proportional transaction costs. This will be
     represented by bid-ask spreads, i.e. shares can be bought at an ask price
     $S^a_t$ or sold at the bid price $S^b_t$, where $S_t^s\geq S^b_t>0,\,
     t=0,1,...,T$ are processes adapted to the filtration $\{\cF_t\}_{t=o}^T$.
     
     The liquidation value at time $t$ of a portfolio $(\gam,\del)$ consisting
     of an amount $\gam$ of cash (or bond) and $\del$ shares of the stock
     equals
     \begin{equation}\label{5.1}
     \te_t(\gam,\del)=\gam+S^b_t\del^+-S^a_t\del^-
     \end{equation}
     which in case $\del<0$ means that a potfolio owner should spend the
     amount $S^a_t\del^-$ in order to close his short position. Observe
     that fractional numbers of shares are allowed here so that both $\gam$
     and $\del$ in a portfolio $(\gam,\del)$ could be, in priciple, any real
     numbers. By definition, a self-financing portfolio strategy is a 
     predictable
     process $(\al_t,\be_t)$ representing positions in cash (or bonds) and
     stock at time $t,\, t=0,1,...,T$ such that 
     \begin{equation}\label{5.2}
     \te_t(\al_t-\al_{t+1},\be_t-\be_{t+1})\geq 0\quad\forall t=0,1,...,T-1
     \end{equation}
     and the set of all such portfolio strategies will be denoted by $\Phi$.

     As before, we consider here a game option which is a contract between its
     seller and buyer such that both have the right to exercise it at any time
     up to a maturity date (horizon) $T$.
     In the presence of transaction costs there
     is a difference whether we stipulate that the option to be settled in cash 
     or both in cash and shares of stock while in the former case an assumption
     concerning transaction costs in the process of portfolio liquidation
     should be made. We adopt here the setup where the payments $X_t$ and $Y_t$
     are made both in cash and in shares of the stock and transaction costs
     take place always when a portfolio adjustment occurs. Thus, the payments
     are, in fact, adapted random 2-vectors $X_t=(X_t^{(1)},X_t^{(2)})$ and
     $Y_t=(Y_t^{(1)},Y_t^{(2)})$ where the first and the second coordinates
     represent, respectively, a cash amount to be payed and a number of stock
     shares to be delivered and as we allow also fractional numbers of shares
     both coordinates can take on any nonnegative real value. The inequality
     $X_t\geq Y_t$ in the zero transaction costs case is replaced in our 
     present setup by
     \begin{equation}\label{5.3}
     \Del_t=\te_t(X^{(1)}_t-Y_t^{(1)},X_t^{(2)}-Y_t^{(2)})\geq 0
     \end{equation}
     and $\Del_t$ is interpreted as a cancellation penalty. We impose also 
     a natural assumption that $X^{(1)}_T=Y^{(1)}_T$ and $X^{(2)}_T=Y^{(2)}_T$,
     i.e. on the maturity date there is no penalty. Therefore, if
     the seller cancells the contract at time $s$ while the buyer exercises
     at time $t$ the former delivers to the latter a package of cash and
     stock shares which can be represented as a 2-vector in the form
     \begin{equation}\label{5.4}
     H(s,t)=(H^{(1)}(s,t),H^{(2)}(s,t))=X_s\bbI_{s<t}+Y_t\bbI_{t\leq s}
     \end{equation}
     where $\bbI_A=1$ if an event $A$ occurs and $\bbI_A=0$ if not. It will
     be convenient to allow the payment components $X_t^{(1)},\, X_t^{(2)}$
     and $Y_t^{(1)},\, Y_t^{(2)}$ to take on any real (and not only
     nonnegative) values which will enable us to demonstrate complete 
     duality (symmetry) between the seller's and the buyer's positions.
     
     A pair $(\sig,\pi)$ of a stopping time $\sig\leq T$ and of a 
     self-financing strategy $\pi=(\al_t,\be_t)^T_{t=0}$ will be called a
     superhedging strategy for the seller of the game option with a payoff
     given by (\ref{5.4}) if for all $t\leq T$,
     \begin{equation}\label{5.5}
     \te_{\sig\wedge t}(\al_{\sig\wedge t}-H^{(1)}(\sig,t),\,
     \be_{\sig\wedge t}-H^{(2)}(\sig,t))\geq 0
     \end{equation}
     where, as usual, $c\wedge d=\min(c,d)$ and $c\vee d=\max(c,d)$. The
     seller's (ask or upper hedging) price $V^a$ of a game option is defined 
     as the infimum of initial amounts required to start a superhedging
     strategy for the seller. Since in order to get $\al_0$ amount of
     cash and $\be_0$ shares of stock at time 0 the seller should spend
     \begin{equation}\label{5.6}
     -\te_0(-\al_0,-\be_0)=\al_0+\be_0^+S_0^a-\be_0^-S^b_0
     \end{equation}
     in cash, we can write
     \begin{eqnarray}\label{5.7}
     &V^a=\inf_{\sig,\pi}\{-\te_0(-\al_0,-\be_0):\, (\sig,\pi)\,\,\mbox{with}\\
     &\pi=(\al_t,\be_t)_{t=0}^T\,\,\mbox{being a superhedging strategy for 
     the seller}\}.\nonumber
     \end{eqnarray}
     
     On the other hand, the buyer may borrow from a bank an amount
     $\te_0(-\al_0,-\be_0)$ to purchase a game option with the payoff
     (\ref{5.4}) and starting with the negative valued portfolio $(\al_0,
     \be_0)$ to manage a self-financing strategy $\pi=(\al_t,\be_t)^T_{t=0}$
     so that for a given stopping time $\tau\leq T$ and all $s\leq T$,
     \begin{equation}\label{5.8}
      \te_{s\wedge\tau}(\al_{s\wedge\tau}+H^{(1)}(s,\tau),\,\be_{s\wedge\tau}
      +H^{(2)}(s,\tau))\geq 0.
     \end{equation}
     In this case the pair $(\tau,\pi)$ will be called a superhedging strategy
     for the buyer. The buyer's (bid or lower hedging) price $V^b$ of the
     game option above is defined as the supremum of initial bank loan 
     required to purchase this game option and to manage a superhedging
     strategy for the buyer. Thus,
     \begin{eqnarray}\label{5.9}
     &V^b=\sup_{\tau,\pi}\{\te_0(-\al_0,-\be_0):\, (\tau,\pi)\,\,\mbox{with}\\
     &\pi=(\al_t,\be_t)_{t=0}^T\,\,\mbox{being a superhedging strategy for 
     the buyer}\}.\nonumber
     \end{eqnarray}
     It follows from the representations of Theorem \ref{thm5.1} below that
     $V^a\geq V^b$.

First, we recall the notion of a randomized stopping time (see \cite{CJ},
\cite{BT}, \cite{RZ} and references there) which is defined as a nonnegative
adapted process $\chi$ such that $\sum_{t=0}^T\chi_t=1$. The set of all 
randomized stopping times will be denoted by $\cX$ while the set of all 
usual or pure stopping times will be denoted by $\cT$. It will be convenient
to identify each pure stopping time $\tau$ with a randomized stopping time
$\chi^\tau$ such that $\chi^\tau_t=\bbI_{\{\tau=t\}}$ for any $t=0,1,...,T$,
so that we could write $\cT\subset\cX$. For any adapted process $Z$ and each
randomized stopping time $\chi$ the time-$\chi$ value of $Z$ is defined by
\begin{equation}\label{5.5.1}
(Z)_\chi=Z_\chi=\sum_{t=0}^T\chi_tZ_t.
\end{equation}

Considering a game option with a payoff given by (\ref{5.4}) we
write also
\begin{equation}\label{5.5.2}
H(\chi,\tilde\chi)=\sum_{s,t=0}^T\chi_s\tilde\chi_tH(s,t)
\end{equation}
which is the seller's payment to the buyer when the former cancells and
the latter exercises at randomized stopping times $\chi$ and $\tilde\chi$,
respectively. In particular, if $\sig$ and $\tau$ are pure stopping times
then
\begin{equation}\label{5.5.3}
H(\chi,\chi^\tau)=\sum_{s=0}^T\chi_sH(s,\tau)\,\,\mbox{and}\,\, H(\chi^\sig,
\chi)=\sum_{t=0}^T\chi_tH(\sig,t).
\end{equation}

Next, we introduce the notion of an approximate martingale which is defined
for any randomized stopping time $\chi$ as a pair $(P,S)$ of a probability
 measure $P$ on $\Om$ and of an adapted process $S$ such that for each
 $t=0,1,...,T$,
 \begin{equation}\label{5.5.7}
 S_t^b\leq S_t\leq S_t^a\,\,\mbox{and}\,\,\chi_{t+1}^*S^b_t\leq\bbE_P(
 S^{\chi^*}_{t+1}|\cF_t)\leq\chi_{t+1}^*S^a_t
 \end{equation}
 where $\bbE_P$ is the expectation with respect to $P$,
 \begin{equation}\label{5.5.8}
 \chi^*_t=\sum_{s=t}^T\chi_s,\, Z_t^{\chi^*}=\sum_{s=t}^T\chi_sZ_s,\,
 \chi^*_{T+1}=0\,\,\mbox{and}\,\, Z_{T+1}^{\chi^*}=0.
 \end{equation}
 Given a randomized stopping time $\chi$ the space of corresponding
  approximate martingales $(P,S)$ will be denoted by $\bar\cP(\chi)$ and
  we denote by $\cP(\chi)$ the subspace of $\bar\cP(\chi)$ consisting of
  pairs $(P,S)$ with $P$ being equivalent to the original (market)
  probability $\bbP$.

Next, we introduce some convex analysis notions and
notations (see \cite{Ro} and \cite{RZ} for more details). Denote by
$\Te$ the family of functions $f:\,\bbR\to\bbR\cup\{-\infty\}$ such that
either $f\equiv -\infty$ or $f$ is a (finite) real valued polyhedral
(continuous piecewise linear with finite number of segments) function. If
$f,g\in\Te$ then, clearly, $f\wedge g,\, f\vee g\in\Te$. The epigraph of
$f\in\Te$ is defined by epi$(f)=\{(x,y)\in\bbR^2:\, x\geq f(y)\}$. For any
$c\geq d$ the function $h_{[d,c]}(y)=cy^--dy^+$, clearly, belongs to $\Te$.
Observe that the self-financing condition (\ref{5.2}) can be rewritten in
the form
\begin{equation}\label{5.5.11}
(\al_t-\al_{t+1},\be_t-\be_{t+1})\in\,\mbox{epi}(h_{[S_t^b,S_t^a]}).
\end{equation}
For each $f\in\Te$ and $c\geq d$ there exists a unique function 
gr$_{[d,c]}(f)\in\Te$ such that
\begin{equation}\label{5.5.12}
\mbox{epi(gr}_{[d,c]}(f))=\mbox{epi}(h_{[d,c]})+\mbox{epi}(f).
\end{equation}

For any $y\in\bbR,\,\mu\in\Om_t$ and $t=0,1,...,T$ define $q_t^a(y)=
q_t^a(\mu,y)$, $q_t^b(y)=q_t^b(\mu,y)$, $r_t^a(y)=r_t^a(\mu,y)$, $r_t^b(y)
=r_t^b(\mu,y)$ by
\begin{eqnarray*}
&q^a_t(y)=X^{(1)}_t+h_{[S^b_t,S^a_t]}(y-X_t^{(2)}),\,\, 
r^a_t(y)=Y^{(1)}_t+h_{[S^b_t,S^a_t]}(y-Y_t^{(2)})\\
&q^b_t(y)=-X^{(1)}_t+h_{[S^b_t,S^a_t]}(y+X_t^{(2)}),\,\, 
r^b_t(y)=-Y^{(1)}_t+h_{[S^b_t,S^a_t]}(y+Y_t^{(2)})
\end{eqnarray*}
with $h_{[d,c]}$ the same as in (\ref{5.5.11}) and (\ref{5.5.12}). Observe that 
if $c\geq d\geq 0$ then either $h_{[d,c]}\equiv 0$ or $h_{[d,c]}$ is a monotone
 decreasing function, and so
\begin{equation}\label{5.5.14}
q^a_t\geq r^a_t\,\,\mbox{and}\,\, q^b_t\leq r^b_t.
\end{equation}
Introduce also
\begin{eqnarray}\label{5.5.15}
&G^{a}_{s,t}(y)=H^{(1)}(s,t)+h_{[S^b_{s\wedge t},S^a_{s\wedge t}]}(y-
H^{(2)}(s,t))=q^a_s(y)\bbI_{s<t}+r^a_t(y)\bbI_{t\leq s}\,\,\mbox{and}\\
&G^{b}_{s,t}(y)=-H^{(1)}(s,t)+h_{[S^b_{s\wedge t},S^a_{s\wedge t}]}(y+
H^{(2)}(s,t))=q^b_s(y)\bbI_{s<t}+r^b_t(y)\bbI_{t\leq s}.\nonumber
\end{eqnarray}
Clearly, the superhedging conditions (\ref{5.5}) of the seller and (\ref{5.8})
of the buyer are equivalent to
\begin{equation}\label{5.5.16}
(\al_{\sig\wedge t},\be_{\sig\wedge t})\in\,\mbox{epi}(G^{a}_{\sig,t})\,\,
\mbox{for all}\,\, t=0,1,...,T\,\,\,\mbox{and}
\end{equation}
\begin{equation}\label{5.5.17}
(\al_{s\wedge\tau},\be_{s\wedge\tau})\in\,\mbox{epi}(G^{b}_{s,\tau})\,\,
\mbox{for all}\,\, s=0,1,...,T,
\end{equation}
respectively. Observe also that
\begin{equation}\label{5.5.18}
 q_t^a(0)=-q^b_t(0)=\te_t(X_t^{(1)},
X_t^{(2)})\,\,\mbox{and}\,\, r_t^a(0)=-r^b_t(0)=\te_t(Y_t^{(1)},Y_t^{(2)}).
\end{equation}
We recall that $X^{(1)}_T=Y^{(1)}_T$ and $X^{(2)}_T=Y^{(2)}_T$, and so 
$q^a_T=r^a_T$ and $q^b_T=r^b_T$. In \cite{Ki6} the following result was
obtained.

  \begin{theorem}\label{thm5.1} {\bf I. Price representations.}
  In the above notations,
  \begin{eqnarray}\label{5.5.9}
  &V^a=\min_{\sig\in\cT}\max_{\chi\in\cX}\max_{(P,S)\in\bar\cP(\chi)}
  \bbE_P\big(H^{(1)}(\sig,\cdot)+SH^{(2)}(\sig,\cdot)\big)_\chi\\
  &=\min_{\sig\in\cT}\max_{\chi\in\cX}\sup_{(P,S)\in\cP(\chi)}
  \bbE_P\big(H^{(1)}(\sig,\cdot)+SH^{(2)}(\sig,\cdot)\big)_\chi\nonumber
  \end{eqnarray}
and
\begin{eqnarray}\label{5.5.10}
  &V^b=\max_{\tau\in\cT}\min_{\chi\in\cX}\min_{(P,S)\in\bar\cP(\chi)}
  \bbE_P\big(H^{(1)}(\cdot,\tau)+SH^{(2)}(\cdot,\tau)\big)_\chi\\
  &=\max_{\tau\in\cT}\min_{\chi\in\cX}\inf_{(P,S)\in\cP(\chi)}
 \bbE_P\big(H^{(1)}(\cdot,\tau)+SH^{(2)}(\cdot,\tau)\big)_\chi\nonumber
  \end{eqnarray}
  where $H^{(1)}(\sig,\cdot),\, H^{(2)}(\sig,\cdot)$ and
   $H^{(1)}(\cdot,\tau),\, H^{(2)}(\cdot,\tau)$ denote functions on
   $\{ 0,1,...,T\}$ whose values at $t$ are obtained by replacing $\cdot$
   by $t$. 

 {\bf II. Recurrent price computations.}

(i) For any $x\in\bbR$, $\mu\in\Om_T$ and 
$\sig\in\cT$ define
\begin{equation}\label{5.5.19}
z^\mu_T(x)=w_T^\mu(x)=r^a_T(\mu,x).
\end{equation}
Next, for $t=1,2,...,T$ and each $\mu\in\Om_{t-1}$ define by backward
induction
\begin{eqnarray}\label{5.5.20}
&\bfz^\mu_{t-1}=\max_{\nu\subset\mu,\,\nu\in\Om_t}z_t^\nu,\,
w^\mu_{t-1}=\mbox{gr}_{[S^b_{t-1}(\mu),S^a_{t-1}(\mu)]}(\bfz^\mu_{t-1}),
\nonumber\\
&z^\mu_{t-1}(x)=\min\big(q^a_{t-1}(\mu,x),\max(r^a_{t-1}(\mu,x),w^\mu_{t-1}(x))
\big).
\nonumber\end{eqnarray}
Then $z_0(0)=V^a$.

(ii) For any $x\in\bbR$, $\mu\in\Om_T$ and $\tau\in\cT$ define
\begin{equation}\label{5.5.21}
u^\mu_T(x)=v^\mu_T(x)=r^b_T(\mu,x).
\end{equation}
Next, for $t=1,2,...,T$ and each $\mu\in\Om_{t-1}$ define by the backward
 induction
 \begin{eqnarray}\label{5.5.22}
&\bfu^\mu_{t-1}=\max_{\nu\subset\mu,\,\nu\in\Om_t}u_t^\nu,\,
v^\mu_{t-1}=\mbox{gr}_{[S^b_{t-1}(\mu),S^a_{t-1}(\mu)]}(\bfu^\mu_{t-1}),\,\,
\nonumber\\
&u^\mu_{t-1}(x)=\min\big(r^b_{t-1}(\mu,x),\max(q^b_{t-1}(\mu,x),v^\mu_{t-1}(x))
\big).\nonumber\\
\nonumber\end{eqnarray}
Then $u_0(0)=-V^b$.

{\bf III. Superhedging strategies.}

(i) Construct by induction a sequence of (pure) 
stopping times $\sig_t\in\cT$ and a self-financing strategy $(\al,\be)$ such
that
\begin{equation}\label{5.5.23}
(\al_t,\be_t)\in\mbox{epi}(z_t)\setminus\mbox{epi}(q^a_t)\,\,\mbox{on}\,\,
\{t<\sig_t\}
\end{equation}
for each $t=0,1,...,T$ in the following way. First, take any 
$\cF_0$-measurable portfolio $(\al_0,\be_0)\in\mbox{epi}(z_0)$ and set
\begin{equation*}
   \sig_0=\left\{\begin{array}{ll}
  0 &\mbox{if}\,\, (\al_0,\be_0)\in\mbox{epi}(q_0^a)\\
  T &\mbox{if}\,\, (\al_0,\be_0)\notin\mbox{epi}(q_0^a).
  \end{array}\right.
  \end{equation*}
  Suppose that an $\cF_t$-measurable portfolio $(\al_t,\be_t)\in
  \mbox{epi}(z_t)$ and a stopping time $\sig_t\in\cT$ have already been
   constructed for some $t=0,1,...T-1$ so that (\ref{5.5.23}) holds true. 
   By (\ref{5.5.12}) and (\ref{5.5.20}),
   \[
   (\al_t,\be_t)\in\mbox{epi}(w_t)=\mbox{epi}(h_{[S^b_t,S^a_t]})+\mbox{epi}
   (\bfz_t)\,\,\mbox{on}\,\,\{t<\sig_t\},
   \]
   and so there exists an $\cF_t$-measurable portfolio $(\al_{t+1},\be_{t+1})$
   such that
   \[
   (\al_{t+1},\be_{t+1})\in\mbox{epi}(\bfz_t),
   \,\, (\al_t-\al_{t+1},\be_t-\be_{t+1})\in\mbox{epi}(h_{[S^b_t,S^a_t]})\,\,
   \mbox{on}\,\,\{t<\sig_t\}
   \]
   and $(\al_{t+1},\be_{t+1})=(\al_t,\be_t)$ on $\{ t\geq\sig_t\}$ which
   provides the self-financing condition (\ref{5.5.11}) both on $\{ t<\sig_t\}$
   and on $\{ t\geq\sig_t\}$. By (\ref{5.5.20}) it follows also that $(\al_{t+1},
   \be_{t+1})\in\,\mbox{epi}(z_{t+1})$ on $\{ t<\sig_t\}\supset
   \{ t+1<\sig_{t+1}\}$. Set
   \begin{equation*}
   \sig_{t+1}=\left\{\begin{array}{ll}
  \sig_t &\mbox{if}\,\, t\geq \sig_t\\
  t+1 &\mbox{if}\,\, t<\sig_t\,\,\mbox{and}\,\,(\al_{t+1},\be_{t+1})\in
  \mbox{epi}(q_{t+1}^a)\\
  T &\mbox{if}\,\, t<\sig_t\,\,\mbox{and}\,\,(\al_{t+1},\be_{t+1})\notin
  \mbox{epi}(q_{t+1}^a).
  \end{array}\right.
  \end{equation*}
  Finally, set $\sig=\sig_T\in\cT$. Then the pair $(\sig,\pi)$ with
  $\pi=(\al,\be)$ constructed by the above algorithm with $(\al_0,\be_0)=
  (V^a,0)$ is a superhedging strategy for the seller.  
  
  (ii) Construct by induction a sequence of (pure) 
stopping times $\tau_t\in\cT$ and a self-financing strategy $(\al,\be)$ such
that
\begin{equation}\label{5.5.24}
(\al_t,\be_t)\in\mbox{epi}(u_t)\setminus\mbox{epi}(r^b_t)\,\,\mbox{on}\,\,
\{t<\tau_t\}
\end{equation}
for each $t=0,1,...,T$ in the following way. First, take any 
$\cF_0$-measurable portfolio $(\al_0,\be_0)\in\mbox{epi}(u_0)$ and set
\begin{equation*}
   \tau_0=\left\{\begin{array}{ll}
  0 &\mbox{if}\,\, (\al_0,\be_0)\in\mbox{epi}(r_0^b)\\
  T &\mbox{if}\,\, (\al_0,\be_0)\notin\mbox{epi}(r_0^b).
  \end{array}\right.
  \end{equation*}
  Suppose that an $\cF_t$-measurable portfolio $(\al_t,\be_t)\in
  \mbox{epi}(u_t)$ and a stopping time $\tau_t\in\cT$ have already been
   constructed for some $t=0,1,...T-1$ so that (\ref{5.5.23}) holds true. 
   By (\ref{5.5.12}) and (\ref{5.5.22}),
   \[
   (\al_t,\be_t)\in\mbox{epi}(v_t)=\mbox{epi}(h_{[S^b_t,S^a_t]})+\mbox{epi}
   (\bfu_t)\,\,\mbox{on}\,\,\{t<\tau_t\},
   \]
   and so there exists an $\cF_t$-measurable portfolio $(\al_{t+1},\be_{t+1})$
   such that
   \[
   (\al_{t+1},\be_{t+1})\in\mbox{epi}(\bfu_t),
   \,\, (\al_t-\al_{t+1},\be_t-\be_{t+1})\in\mbox{epi}(h_{[S^b_t,S^a_t]})\,\,
   \mbox{on}\,\,\{t<\tau_t\}
   \]
   and $(\al_{t+1},\be_{t+1})=(\al_t,\be_t)$ on $\{ t\geq\tau_t\}$ which
   provides the self-financing condition (\ref{5.5.11}) both on $\{ t<\tau_t\}$
   and on $\{ t\geq\tau_t\}$. By (\ref{5.5.22}) it follows also that $(\al_{t+1},
   \be_{t+1})\in\,\mbox{epi}(u_{t+1})$ on $\{ t<\tau_t\}\supset
   \{ t+1<\tau_{t+1}\}$. Set
   \begin{equation*}
   \tau_{t+1}=\left\{\begin{array}{ll}
  \tau_t &\mbox{if}\,\, t\geq \tau_t\\
  t+1 &\mbox{if}\,\, t<\tau_t\,\,\mbox{and}\,\,(\al_{t+1},\be_{t+1})\in
  \mbox{epi}(r_{t+1}^b)\\
  T &\mbox{if}\,\, t<\tau_t\,\,\mbox{and}\,\,(\al_{t+1},\be_{t+1})\notin
  \mbox{epi}(r_{t+1}^b).
  \end{array}\right.
  \end{equation*}
  Finally, set $\tau=\tau_T\in\cT$. Then the pair $(\tau,\pi)$ with
  $\pi=(\al,\be)$ constructed by the above algorithm with $(\al_0,\be_0)=
  (-V^b,0)$ is a superhedging strategy for the buyer.
   \end{theorem}
   
  There are by now very few papers on game options with transaction costs.
  In \cite{Do2} it is shown that the cheapest superhedging strategy for
  a game option in a Black-Scholes market with transaction costs is 
  the buy-and-hold portfolio strategy together with a hitting time of a
  Borel set. The shortfall risk for a game option in a Black-Scholes market 
  with transaction costs is obtained in \cite{Do3} as a limit of corresponding
  expressions for a sequence of binomial models in the spirit of Section 
  \ref{sec4} above.

\end{document}